\documentclass[twocolumn]{aastex62}
\usepackage{txfonts, lineno, subfigure}
\usepackage{subfigure}
\shorttitle{Search for large-scale anisotropy on arrival directions of UHECRs observed with TA}
\shortauthors{Published in ApJL, 898, L28 (2020), https://doi.org/10.3847/2041-8213/aba0bc}
\begin{document}

\title{Search for Large-scale Anisotropy on Arrival Directions of Ultra-high-energy Cosmic Rays Observed with the Telescope Array Experiment}

\correspondingauthor{Toshihiro Fujii}
\email{fujii@cr.scphys.kyoto-u.ac.jp}

\collaboration{Telescope Array Collaboration}

\author[0000-0001-6141-4205]{R.U. Abbasi}
\affiliation{Department of Physics, Loyola University Chicago, Chicago, Illinois, USA}

\author{M. Abe}
\affiliation{The Graduate School of Science and Engineering, Saitama University, Saitama, Saitama, Japan}

\author[0000-0001-5206-4223]{T. Abu-Zayyad}
\affiliation{High Energy Astrophysics Institute and Department of Physics and Astronomy, University of Utah, Salt Lake City, Utah, USA}

\author{M. Allen}
\affiliation{High Energy Astrophysics Institute and Department of Physics and Astronomy, University of Utah, Salt Lake City, Utah, USA}

\author{R. Azuma}
\affiliation{Graduate School of Science and Engineering, Tokyo Institute of Technology, Meguro, Tokyo, Japan}

\author{E. Barcikowski}
\affiliation{High Energy Astrophysics Institute and Department of Physics and Astronomy, University of Utah, Salt Lake City, Utah, USA}

\author{J.W. Belz}
\affiliation{High Energy Astrophysics Institute and Department of Physics and Astronomy, University of Utah, Salt Lake City, Utah, USA}

\author{D.R. Bergman}
\affiliation{High Energy Astrophysics Institute and Department of Physics and Astronomy, University of Utah, Salt Lake City, Utah, USA}

\author{S.A. Blake}
\affiliation{High Energy Astrophysics Institute and Department of Physics and Astronomy, University of Utah, Salt Lake City, Utah, USA}

\author{R. Cady}
\affiliation{High Energy Astrophysics Institute and Department of Physics and Astronomy, University of Utah, Salt Lake City, Utah, USA}

% ...

\begin{abstract}
Motivated by the detection of a significant dipole structure in the arrival directions of ultra-high-energy cosmic rays above 8\,EeV reported by the Pierre Auger Observatory (Auger), 
we search for a large-scale anisotropy using data collected with the surface detector array of the Telescope Array Experiment (TA).
With 11 years of TA data, a dipole structure in a projection of the right ascension is fitted with an amplitude of 3.3$\pm$1.9\% and a phase of $131^{\circ}\pm33^{\circ}$.
The corresponding 99\% confidence-level upper limit on the amplitude is 7.3\%.
At the current level of statistics, the fitted result is compatible with both an isotropic distribution and the dipole structure reported by Auger.
\end{abstract}



\section{Introduction}

Clarifying the origin and nature of the ultra-high-energy cosmic rays (UHECRs) has been a decades-long endeavor~\citep{Dawson:2017rsp,AlvesBatista:2019tlv}.
There are a number of challenges in identifying sources, including the uncertainty in the chemical composition of cosmic rays with energies above 10\,EeV ($\equiv 10^{19}$\,eV), and existence of galactic and extragalactic magnetic fields that scramble directional information.

The origin of UHECRs above 50\,EeV would be significantly restricted to nearby sources due to interactions with the cosmic microwave background radiation via pion production for protons or via photo-disintegration processes for heavier nuclei, known as the GZK cutoff~\citep{bib:gzk1,bib:gzk2}.
With small deflections of UHECRs by galactic and intergalactic magnetic fields, $\sim5^{\circ} Z (E/50\,\rm{EeV})^{-1}$ where $Z$ is the charge of nuclei~\citep{Bray:2018ipq}, a small-intermediate scale anisotropy
\footnote{In this research field, we define a small angular scale anisotropy as less than 5$^{\circ}$, intermediate scale as 10$^{\circ}$ to 35$^{\circ}$ and large scale as $>40^{\circ}$.}
is predicted at the highest energies.
On the other hand, the distance to sources and deflections by magnetic fields are enlarged at lower energies around 10\,EeV, resulting in a large-scale anisotropy, which can be approximated as a dipole~\citep{Harari:2015hba,diMatteo:2017dtg,Eichmann:2020adn}.
Full-sky anisotropy measurements from both hemispheres are also essential to understand the large-scale anisotropy.

The two largest observatories, the Telescope Array Experiment (TA)~\citep{Tokuno:2012mi,AbuZayyad:2012kk} and Pierre Auger Observatory (Auger)~\citep{ThePierreAuger:2015rma}, are currently in operation and observing UHECRs from
the Northern and Southern hemispheres, respectively.
There are indications of a number of intriguing intermediate-scale
anisotropies at the highest energies ($E > 30$\,EeV), such as the hotspot reported by TA~\citep{Abbasi:2014lda} and the result of the flux pattern correlation analysis from Auger~\citep{Aab:2018chp} and TA~\citep{Abbasi:2018tqo}. 
However, no conclusive results are reported and the origins of UHECRs are still unknown.

In 2017 the Pierre Auger Collaboration reported the observation of a significant large-scale anisotropy in the arrival directions of cosmic rays above 8\,EeV, 
indicating an obvious dipole structure of 4.7\% amplitude in a projection of the right ascension with a 5.2$\sigma$ significance~\citep{Aab:2017tyv}.
An enhancement of the dipole amplitude above 4\,EeV and results down to 0.03\,EeV are also reported~\citep{Aab:2018mmi,Aab:2020xgf}
These results are consistent with an extragalactic origin of UHECRs.
The dipole anisotropy of right ascension over a broad energy range is reported by a variety of experiments, as summarized in~\citep{Mollerach:2017idb}. 

In this letter we investigate the existence of a dipole structure using TA data. 
TA detects UHECRs from the Northern hemisphere, 
providing us with an independent check of the large-scale anisotropy reported by Auger.

\section{Telescope Array Experiment}
TA is the largest cosmic-ray detector in the Northern hemisphere, located near the city of Delta, Utah, USA (39.30$^{\circ}$ North and 112.91$^{\circ}$ West at $\sim$1400\,m above sea level)~\citep{AbuZayyad:2012kk}.
The surface detector array (SD) consists of 507 plastic scintillators of 3\,m$^2$ area deployed in a square grid with a 1.2\,km spacing.
The total effective area is approximately 700\,km$^2$. Additional surface detectors, designed to provide a fourfold increase in effective area, named TAx4, are now being deployed and partly in operation~\citep{Kido:2019enj}.
The TA SD is overlooked by fluorescence detectors, which are used for determination of the calorimetric energy of an air shower from energy deposited in the atmosphere during its development~\citep{Tokuno:2012mi}. 

The arrival direction of UHECR as measured by the TA SD is determined from the relative difference in arrival time of the shower front at each surface detector (which are time-synchronized using GPS modules).
The energy estimator of the TA SD is the particle density measured at a distance of 800\,m from the air shower axis, called $S_{800}$.
The $S_{800}$ parameter is converted to the primary energy as a function of zenith angle based on a Monte Carlo simulation using the CORSIKA software package~\citep{Heck:1998vt}.
The obtained energy is calibrated to the calorimetric energy measured by the fluorescence detectors using a scaling factor of 1/1.27~\citep{AbuZayyad:2012ru}.
The typical resolution of the TA SD is 1.0$^{\circ}$ $\sim$ 1.5$^{\circ}$ in arrival direction and 10\% $\sim$ 15\% in primary energy~\citep{AbuZayyad:2012ru}, 
and the systematic uncertainty in the energy scale is quoted as 21\%~\citep{TheTelescopeArray:2015mgw}.

\section{Data-set and methodology}
TA SD data recorded over 11 years from May 2008 to May 2019 were analyzed for a study of the large-scale anisotropy.
To avoid potential penalties from scanning, we use an \textit{a priori} energy threshold of 8.8\,EeV, equivalent to 8\,EeV used by Auger, taking into account the 10\% energy scale difference between TA and Auger
calibrated from a common hardening feature of both energy spectra around 5\,EeV~\citep{Verzi:2017hro}.
There were 6032 events surviving this cut above a primary energy of 8.8\,EeV, with zenith angles below 55$^{\circ}$ and the same quality cuts used in the TA spectrum analysis~\citep{AbuZayyad:2012ru}.
In this data-set, TA SD is capable of measuring UHECRs injected in a declination band from $-15^{\circ}$ to $90^{\circ}$.

As the trigger efficiency of the TA SD at 8.8\,EeV is not 100\%, the obtained right ascension distribution is compared with the expected distribution assuming an isotropic UHECR sky from a time-dependent Monte Carlo (MC) simulation including calibration constants, live time and dead time of each surface detector station, and TA SD trigger efficiencies.
A sidereal time distribution (366.25 cycles/year) of the simulation has a small amplitude of $0.4\%$.
The residual intensity is defined as $(N_{\rm{obs}} - N_{\rm{exp}})/N_{\rm{exp}}$, where $N_{\rm{obs}}$ is the number of events observed by TA and $N_{\rm{exp}}$ is the expected number estimated from the simulation.
The number of total events in the MC simulation is normalized to the number of events in the TA data above 8.8\,EeV.

We also check for possible declination band dependence in common with Auger with $\delta < 24.8^{\circ}$ where $\delta$ is the declination, and a higher declination band with $\delta \ge 24.8^{\circ}$.
These declination bands were defined in~\citep{Verzi:2017hro}.

\section{Results}
\subsection{Dipole structure in right ascension}

\begin{figure}
  \includegraphics[width=1.0\linewidth]{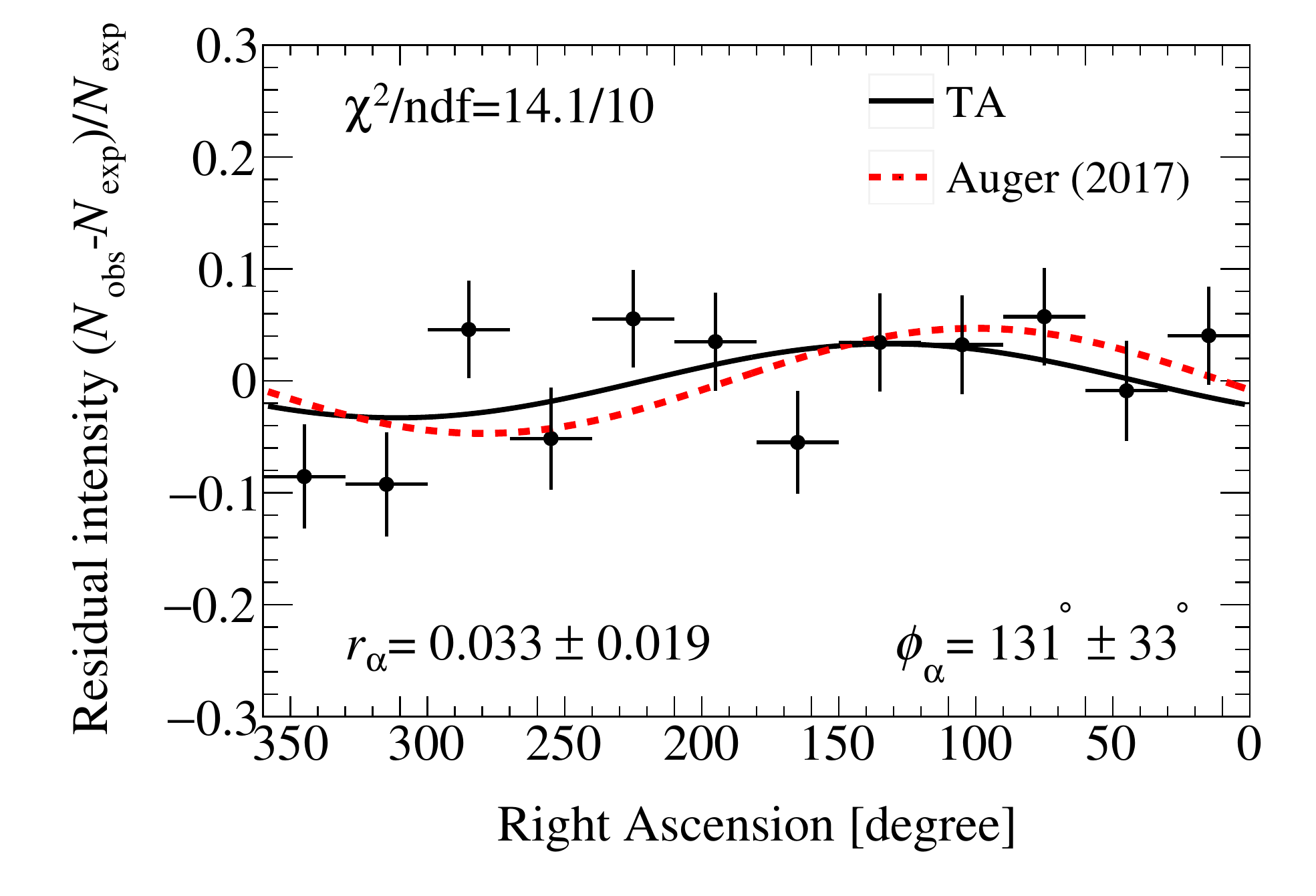}
  \caption{Residual intensities of UHECRs with energies above 8.8\,EeV observed with 11 years of TA data as a function of the right ascension. The black curve is the TA fitted dipole result and the red dashed curve is the Auger reported dipole result.}
  \label{fig:dipole}
\end{figure}

\begin{figure}
  \subfigure[Equatorial coordinates]{\includegraphics[width=1\linewidth]{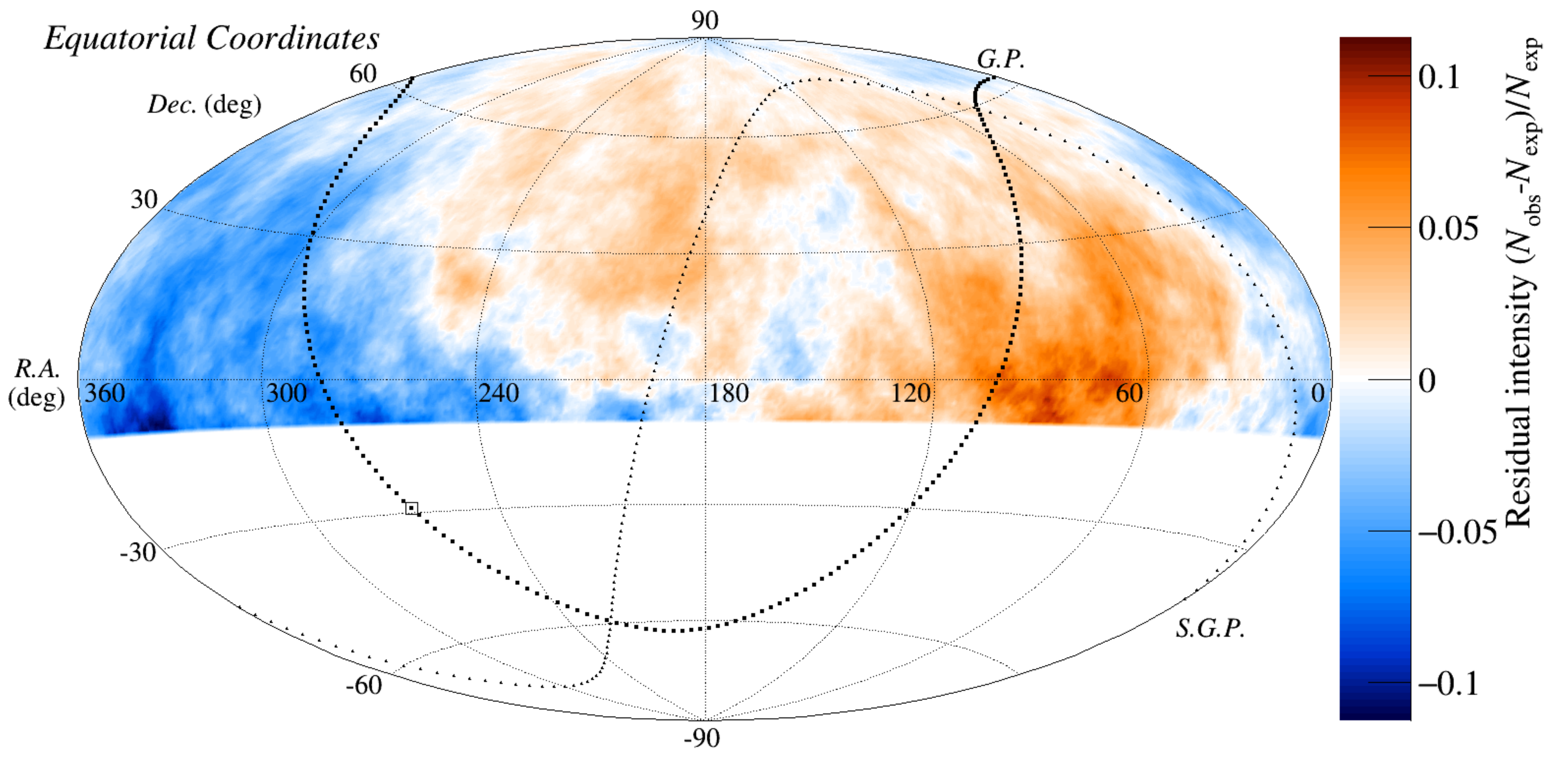}}
  \subfigure[Galactic coordinates]{\includegraphics[width=1\linewidth]{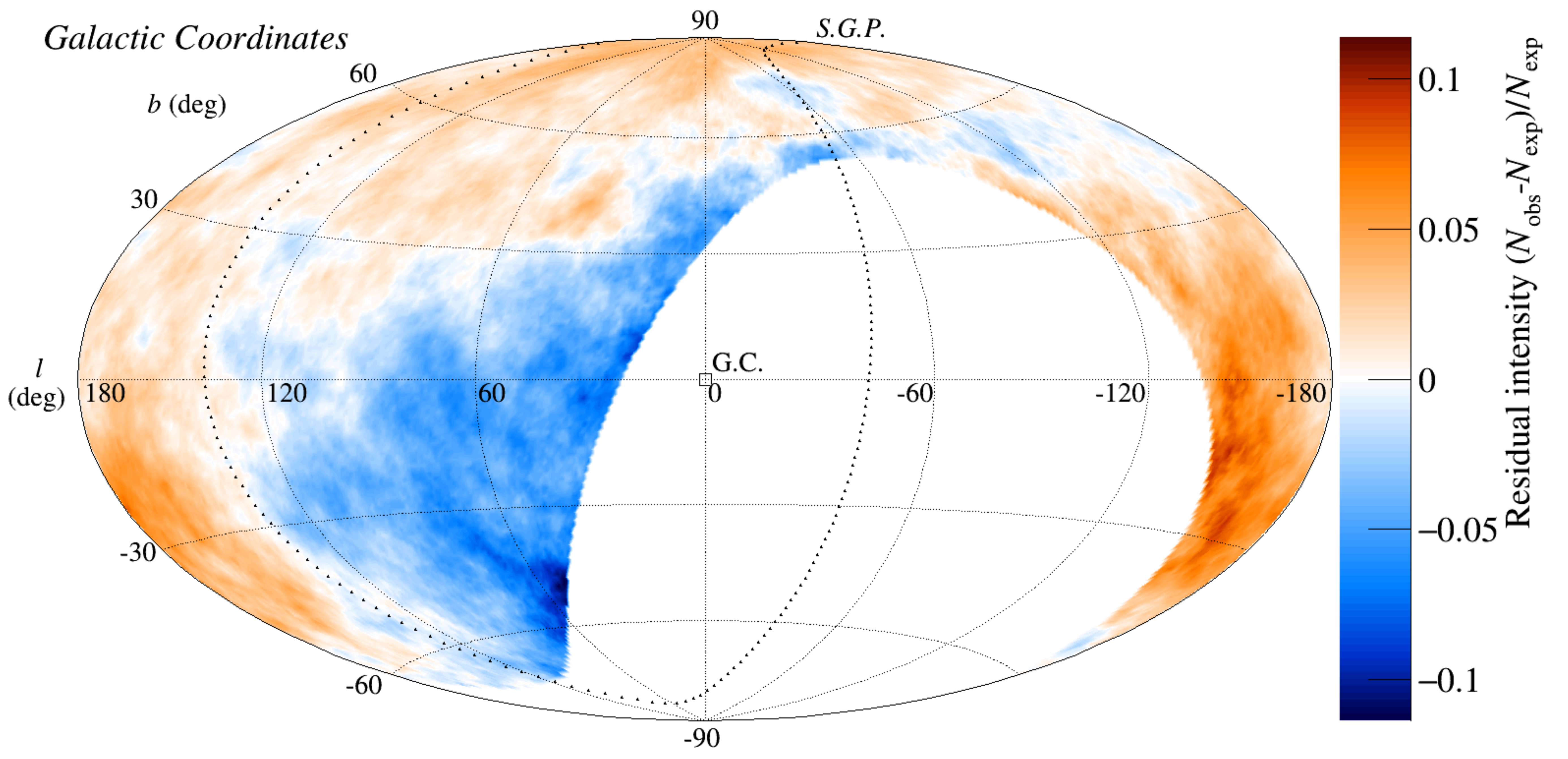}}
  \caption{Sky map of residual intensities between TA data and an isotropic distribution shown in the Equatorial and the Galactic coordinates. The arrival directions are oversampled with a 45$^{\circ}$ radius cylindrical function. The galactic plane (G.P.) and the super-galactic plane (S.G.P.) are shown as thick and thin dotted curves, respectively. The galactic center (G.C.) is indicated by the open square.}
  \label{fig:skymap}
\end{figure}

\begin{figure}
 \subfigure[Higher declination band, $\delta \ge 24.8^{\circ}$]{\includegraphics[width=1.0\linewidth]{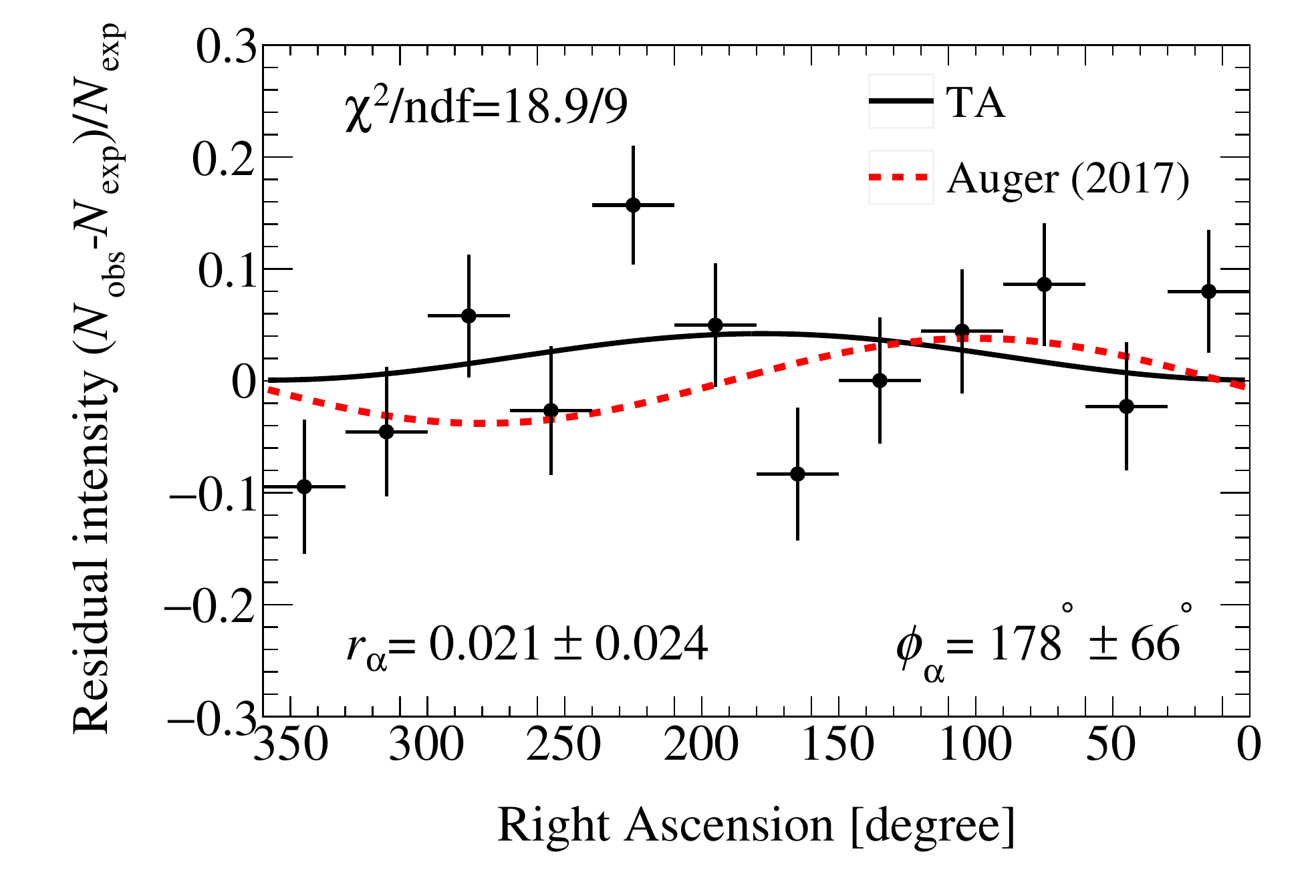}}
 \subfigure[Common declination band with Auger, $\delta < 24.8^{\circ}$]{\includegraphics[width=1.0\linewidth]{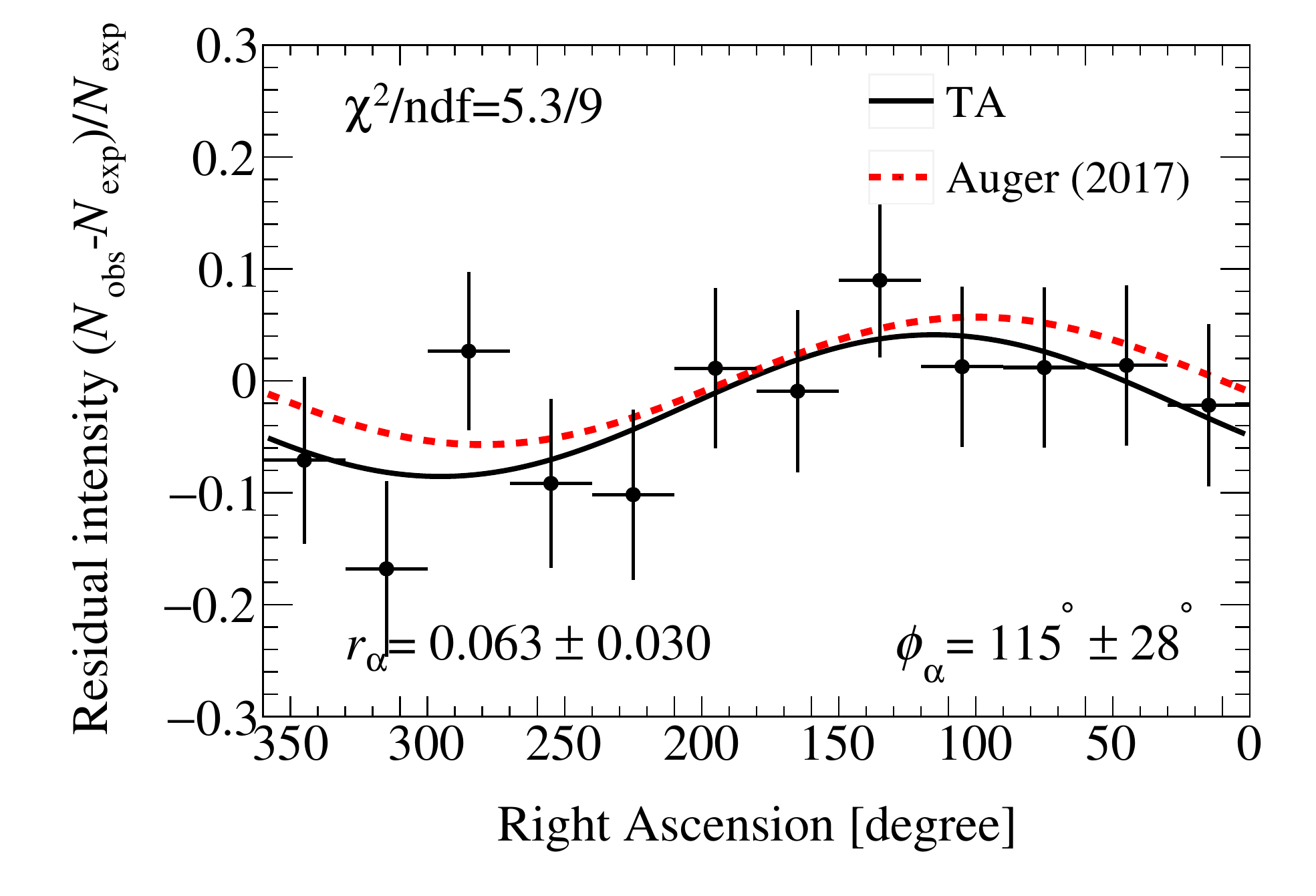}}
 \caption{Residual intensity in right ascension in (a) the higher declination band, $\delta \ge 24.8^{\circ}$, and (b) the common declination band with Auger, $\delta < 24.8^{\circ}$. The solid black curve is the fit to the TA SD data, and the red dashed curve is the expected dipole structure from the Auger result, 3.8\% amplitude in the higher band and 5.7\% amplitude in the common band.} 
 \label{fig:dec}
\end{figure}

Figure~\ref{fig:dipole} shows the residual intensity as a function of right ascension between 11 years of TA SD data and the isotropic expectation calculated from the MC simulation. 
The residual intensity is fitted to $r_{\alpha} \cos(x - \phi_{\alpha})$, where $r_{\alpha}$ is the amplitude of the dipole and $\phi_{\alpha}$ is the phase. 
The obtained dipole structure has an amplitude of 3.3$\pm$1.9\% with a phase of $131^{\circ}\pm33^{\circ}$. 
The TA SD data points are also compared with an isotropic distribution and the dipole structure reported from Auger, which has a 4.7\% amplitude with a phase of 100$^{\circ}$. 
The $\chi^{2}$/ndf against the isotropic distribution is 17.1/12 with a probability of 14\%, and that against the Auger dipole is 15.9/12 with a probability of 20\%.
With current statistics, the obtained TA SD result is consistent with both hypotheses.

Since there is no significant dipole structure, an upper limit is evaluated for a chance probability of 1\% that a fluctuation of an isotropic source distribution would yield an amplitude greater than the measured value. 
The obtained 99\% confidence-level upper-limit on amplitude is 7.3\% for TA data above 8.8\,EeV.

\subsection{Systematic uncertainty}


When we change to an assumption of 100\% trigger efficiency and of uniform exposure to calculate the expected number of events from each right ascension bin, we have a dipole shifted by the amplitude of $-0.4\%$ and the phase of $+5^{\circ}$. This gives a possible spurious dipole due to time-dependent detector conditions.

A modulation related to the atmospheric condition is seen in the TA SD data.
When an empirical energy correction as a function of the atmospheric density at the ground is applied, the fitted dipole result is changed by $-1.1\%$ in amplitude and $-23^{\circ}$ in phase. The correction method related to the atmospheric condition is being investigated and validated.
These systematic uncertainties are smaller than the current statistical uncertainty of $1.9\%$ in amplitude in the TA SD data.

The anti-sidereal time distribution (364.25 cycles/year) is also tested for a systematic study. The fitted amplitude is 1.8$\pm$1.8\%, being consistent with a uniform distribution.

\subsection{Residual intensity sky map}
To investigate the dipole structure, the residual-intensity sky map is calculated with a 45$^{\circ}$ radius cylindrical smoothing function, usually known as a top-hat function, applied to the arrival directions of UHECRs as used in Auger~\citep{Aab:2017tyv}.
Figure~\ref{fig:skymap} shows the residual-intensity sky map of UHECRs measured by TA with energies above 8.8\,EeV in equatorial and galactic coordinates.
Even with our limited statistics, a similar dipole structure is seen in the common declination  $\delta < 24.8^{\circ}$ band shared with Auger.

\subsection{Declination dependence}

Figure~\ref{fig:dec} shows the residual intensities and fitted results in the higher declination band above 24.8$^{\circ}$ and the declination band in common with Auger below 24.8$^{\circ}$.
If we assume the three-dimensional dipole structure reported by Auger,  with 6.5\% amplitude at $100^{\circ}$ in right ascension and $-24^{\circ}$ in declination~\citep{Aab:2017tyv}, the expected amplitudes are evaluated to be 3.8\% in the higher declination band and 5.7\% in the common declination band with Auger.
These expectations are indicated by the curves in the plots of Figure~\ref{fig:dec}.
An offset parameter is added to the fitting function as we divide the data into two sets without a re-normalization in each dataset.

The obtained results are still inconclusive.
The fitted function in the higher declination band has an amplitude of $2.1\pm2.4$\% at $178^{\circ}\pm66^{\circ}$.
The $\chi^{2}$/ndf against the isotropic distribution is 21.2/12 with a probability of 5\%, and that against the reported dipole from Auger is 23.1/12 with a probability of 3\%.
The dipole structure in the common declination band with Auger is fitted by an amplitude of $6.3\pm3.0\%$ at $115^{\circ}\pm28^{\circ}$, which is consistent with the one reported by Auger.
Also the $\chi^{2}$/ndf against the isotropic distribution is 10.8/12 with a probability of 55\%, and that against the Auger result is 6.8/12 with a probability of 80\%.
The fitted parameters and corresponding upper limits at 99\% confidence level in these bands are summarized in Table~\ref{tab:dipole},

\begin{table}
  \centering
  \begin{tabular}{l|ccccc}
                          & $E_{\rm{med}}$ [EeV] &  $N$   & $r_{\alpha}$ [\%] & $\phi_{\alpha}$ [$^{\circ}$] & $r_{\alpha}^{\rm{UL}}$ [\%]  \\ \hline
    All                           & 13.0 &  6032  & 3.3$\pm$1.9  & $131\pm33$  &  7.3  \\
    $\delta$ $\ge$ 24.8$^{\circ}$ &  -   &  3778  & 2.1$\pm$2.4  & $178\pm66$  &  6.7  \\
    $\delta$ $<$ 24.8$^{\circ}$   &  -   &  2254  & 6.3$\pm$3.0  & $115\pm28$  & 12.9  \\
  \end{tabular}
  \caption{Summary of the median energy, number of events, the dipole fitted results and the corresponding upper limits (UL) amplitude at 99\% confidence level in all bands, in the higher declination band, $\delta$ $\ge$ 24.8$^{\circ}$, and in the common declination band with Auger, $\delta$ $<$ 24.8$^{\circ}$.}
  \label{tab:dipole}
\end{table}

\section{Discussion}
Due to the limited statistics with the TA data, the obtained amplitudes are compatible with an isotropic distribution at a $2\sigma$ significance level.
However, focusing on the common declination band with Auger and the sky map of residual intensities in Figure~\ref{fig:skymap}, the fitted phase is apart from the galactic center of $266^{\circ}$ in right ascension, supporting an extragalactic origin of UHECRs reported from Auger.
In the higher declination band, probabilities of 5\% against the isotropic distribution and of 3\% against the Auger reported dipole might be influenced by additional components of the anisotropy, possibly related to local sources and magnetic fields.

Further data-taking with TA and TA$\times$4 will be essential to differentiate between the isotropic and the dipole structure hypotheses, and capable to investigate an energy dependence of the dipole amplitude reported from Auger~\citep{Aab:2018mmi}. 
Continuing the full-sky anisotropy searches from both hemispheres is of the upmost importance to clarify the origin and nature of UHECRs.


\section{Conclusions}
We report on a follow-up search for the dipole structure reported by Auger using 11 years of TA SD data from the Northern sky.
We see results consistent with both an isotropic source distribution and the dipole structure reported by Auger.
Therefore we have evaluated a 99\% confidence-level upper limit of $r_{\alpha}^{\rm{UL}} = 7.3\%$ above 8.8\,EeV on the amplitude of a dipole structure in a projection of the right ascension.
Although the residual intensity sky map shows a similar dipole structure to Auger, further statistics from TA are required to distinguish the two hypotheses.
Further data collection by TA and the on-going upgrade of TA$\times$4 will be essential for further studies.

\section*{Acknowledgments}

The Telescope Array experiment is supported by the Japan Society for
the Promotion of Science(JSPS) through
Grants-in-Aid
for Priority Area
431,
for Specially Promoted Research
JP21000002,
for Scientific  Research (S)
JP19104006,
for Specially Promoted Research
JP15H05693,
for Scientific  Research (S)
JP15H05741, for Science Research (A) JP18H03705,
for Young Scientists (A)
JPH26707011,
and for Fostering Joint International Research (B)
JP19KK0074,
by the joint research program of the Institute for Cosmic Ray Research (ICRR), The University of Tokyo;
by the U.S. National Science
Foundation awards PHY-0601915,
PHY-1404495, PHY-1404502, and PHY-1607727;
by the National Research Foundation of Korea
(2016R1A2B4014967, 2016R1A5A1013277, 2017K1A4A3015188, 2017R1A2A1A05071429) ;
by the Russian Academy of
Sciences, RFBR grant 20-02-00625a (INR), IISN project No. 4.4502.13, and Belgian Science Policy under IUAP VII/37 (ULB). The foundations of Dr. Ezekiel R. and Edna Wattis Dumke, Willard L. Eccles, and George S. and Dolores Dor\'e Eccles all helped with generous donations. The State of Utah supported the project through its Economic Development Board, and the University of Utah through the Office of the Vice President for Research. The experimental site became available through the cooperation of the Utah School and Institutional Trust Lands Administration (SITLA), U.S. Bureau of Land Management (BLM), and the U.S. Air Force. We appreciate the assistance of the State of Utah and Fillmore offices of the BLM in crafting the Plan of Development for the site.  Patrick Shea assisted the collaboration with valuable advice  on a variety of topics. The people and the officials of Millard County, Utah have been a source of steadfast and warm support for our work which we greatly appreciate. We are indebted to the Millard County Road Department for their efforts to maintain and clear the roads which get us to our sites. We gratefully acknowledge the contribution from the technical staffs of our home institutions. An allocation of computer time from the Center for High Performance Computing at the University of Utah is gratefully acknowledged.

T.F. greatly appreciate the support of the Hakubi Center for Advanced Research, Kyoto University.

\bibliographystyle{aasjournal}
\bibliography{main}

\begin{thebibliography}{}
\expandafter\ifx\csname natexlab\endcsname\relax\def\natexlab#1{#1}\fi
\providecommand{\url}[1]{\href{#1}{#1}}
\providecommand{\dodoi}[1]{doi:~\href{http://doi.org/#1}{\nolinkurl{#1}}}
\providecommand{\doeprint}[1]{\href{http://ascl.net/#1}{\nolinkurl{http://ascl.net/#1}}}
\providecommand{\doarXiv}[1]{\href{https://arxiv.org/abs/#1}{\nolinkurl{https://arxiv.org/abs/#1}}}

\bibitem[{Aab {et~al.}(2015)}]{ThePierreAuger:2015rma}
Aab, A., {et~al.} 2015, Nucl. Instrum. Meth., A798, 172,
  \dodoi{10.1016/j.nima.2015.06.058}

\bibitem[{Aab {et~al.}(2017)}]{Aab:2017tyv}
---. 2017, Science, 357, 1266, \dodoi{10.1126/science.aan4338}

\bibitem[{Aab {et~al.}(2018{\natexlab{a}})}]{Aab:2018chp}
---. 2018{\natexlab{a}}, Astrophys. J., 853, L29,
  \dodoi{10.3847/2041-8213/aaa66d}

\bibitem[{Aab {et~al.}(2018{\natexlab{b}})}]{Aab:2018mmi}
---. 2018{\natexlab{b}}, Astrophys. J., 868, 4,
  \dodoi{10.3847/1538-4357/aae689}

\bibitem[{Aab {et~al.}(2020)}]{Aab:2020xgf}
---. 2020, Astrophys. J., 891, 142, \dodoi{10.3847/1538-4357/ab7236}

\bibitem[{Abbasi {et~al.}(2014)}]{Abbasi:2014lda}
Abbasi, R.~U., {et~al.} 2014, Astrophys. J., 790, L21,
  \dodoi{10.1088/2041-8205/790/2/L21}

\bibitem[{Abbasi {et~al.}(2016)}]{TheTelescopeArray:2015mgw}
---. 2016, Astropart. Phys., 80, 131,
  \dodoi{10.1016/j.astropartphys.2016.04.002}

\bibitem[{Abbasi {et~al.}(2018)}]{Abbasi:2018tqo}
---. 2018, Astrophys. J., 867, L27, \dodoi{10.3847/2041-8213/aaebf9}

\bibitem[{Abu-Zayyad {et~al.}(2013{\natexlab{a}})}]{AbuZayyad:2012kk}
Abu-Zayyad, T., {et~al.} 2013{\natexlab{a}}, Nucl. Instrum. Meth., A689, 87,
  \dodoi{10.1016/j.nima.2012.05.079}

\bibitem[{Abu-Zayyad {et~al.}(2013{\natexlab{b}})}]{AbuZayyad:2012ru}
---. 2013{\natexlab{b}}, Astrophys. J., 768, L1,
  \dodoi{10.1088/2041-8205/768/1/L1}

\bibitem[{Alves~Batista {et~al.}(2019)}]{AlvesBatista:2019tlv}
Alves~Batista, R., {et~al.} 2019, Front. Astron. Space Sci., 6, 23,
  \dodoi{10.3389/fspas.2019.00023}

\bibitem[{Bray \& Scaife(2018)}]{Bray:2018ipq}
Bray, J., \& Scaife, A. 2018, Astrophys. J., 861, 3,
  \dodoi{10.3847/1538-4357/aac777}

\bibitem[{Dawson {et~al.}(2017)Dawson, Fukushima, \& Sokolsky}]{Dawson:2017rsp}
Dawson, B.~R., Fukushima, M., \& Sokolsky, P. 2017, PTEP, 2017, 12A101,
  \dodoi{10.1093/ptep/ptx054}

\bibitem[{di~Matteo \& Tinyakov(2018)}]{diMatteo:2017dtg}
di~Matteo, A., \& Tinyakov, P. 2018, Mon. Not. Roy. Astron. Soc., 476, 715,
  \dodoi{10.1093/mnras/sty277}

\bibitem[{Eichmann \& Winchen(2020)}]{Eichmann:2020adn}
Eichmann, B., \& Winchen, T. 2020, JCAP, 04, 047,
  \dodoi{10.1088/1475-7516/2020/04/047}

\bibitem[{Greisen(1966)}]{bib:gzk1}
Greisen, K. 1966, Phys.Rev.Lett., 16, 748, \dodoi{10.1103/PhysRevLett.16.748}

\bibitem[{Harari {et~al.}(2015)Harari, Mollerach, \& Roulet}]{Harari:2015hba}
Harari, D., Mollerach, S., \& Roulet, E. 2015, Phys. Rev. D, 92, 063014,
  \dodoi{10.1103/PhysRevD.92.063014}

\bibitem[{Heck {et~al.}(1998)Heck, Knapp, Capdevielle, Schatz, \&
  Thouw}]{Heck:1998vt}
Heck, D., Knapp, J., Capdevielle, J.~N., Schatz, G., \& Thouw, T. 1998,
  Forschungszentrum Karlsruhe Report, FZKA-6019

\bibitem[{Kido {et~al.}(2019)}]{Kido:2019enj}
Kido, E., {et~al.} 2019, EPJ Web Conf., 210, 06001,
  \dodoi{10.1051/epjconf/201921006001}

\bibitem[{Mollerach \& Roulet(2018)}]{Mollerach:2017idb}
Mollerach, S., \& Roulet, E. 2018, Prog. Part. Nucl. Phys., 98, 85,
  \dodoi{10.1016/j.ppnp.2017.10.002}

\bibitem[{Tokuno {et~al.}(2012)}]{Tokuno:2012mi}
Tokuno, H., {et~al.} 2012, Nucl. Instrum. Meth., A676, 54,
  \dodoi{10.1016/j.nima.2012.02.044}

\bibitem[{Verzi {et~al.}(2017)Verzi, Ivanov, \& Tsunesada}]{Verzi:2017hro}
Verzi, V., Ivanov, D., \& Tsunesada, Y. 2017, PTEP, 2017, 12A103,
  \dodoi{10.1093/ptep/ptx082}

\bibitem[{Zatsepin \& Kuzmin(1966)}]{bib:gzk2}
Zatsepin, G., \& Kuzmin, V. 1966, JETP Lett., 4, 78

\end{thebibliography}

\end{document}